# An Innovative Computational Approach for Modeling Thermo-hydro Processes within Enhanced Geothermal System

Kamran Jahan Bakhsh, Masami Nakagawa, Mahmood Arshad, Lucila Dunnington

kjahanab@mines.edu, Mnakagaw@mines.edu, Harshad@mines.edu, ldunning@mines.edu

**ABSTRACT**

An innovative computational approach to capturing the essential aspects of an Enhanced Geothermal System (EGS) is formulated. The modified finite element method is utilized to model transient heat and fluid flow within an EGS reservoir. Three main features are modified to determine their impact on the reservoir: the fracture model, the porous model, and the coupling physical model. For the first feature, the fractures are modeled as a two-dimensional subdomain embedded in a vast three-dimensional rock mass. The fracture model eliminates the need to create slender fractures with a high aspect ratio by allowing reduction of the spatial discretization of the fractures from three- to two-dimensional finite elements. In this model, pseudo three-dimensional equations are adopted to drive the physics of heat and fluid flow in the fractures. In the second feature, a porous subdomain with effective transport properties is modeled to integrate a thermally-shocked region in the reservoir simulation. In the third feature, three-dimensional heat flow in the rock mass is coupled to the two-dimensional heat and fluid flow in the fractures. Numerical examples are computed to illustrate the computational capability of the proposed model to simulate heat and fluid flow in an EGS. Results show that the proposed model is capable of both effectively integrating thermal fractures into reservoir simulation, as well as efficiently tackling the computational burden exerted by high aspect ratio geometries. A parametric analysis is also performed in which the effect of thermal fractures and thermally-shocked region on reservoir performance is evaluated.

**1. INTRODUCTION**

The rapid increase in energy demand in conjunction with climate change mitigation endorses the transition of the energy supply of the world to Renewable Energy (RE) sources. Clean renewable energy sources, i.e., wind, solar and geothermal have huge potential for electricity generation with a low-carbon footprint. Wind and solar are the fastest-growing renewable energy sources (Jessica Shankleman 2016; Sawin et al. 2016), due to their intermittent nature, however, the electricity output of these RE technologies is variable and unpredictable (Edenhofer et al. 2011; Skea et al. 2008). However, geothermal energy is a promising alternative for fossil fuel-based power plants. Although harvesting just 1% of the Earth's crust heat would supply the world's current energy demand for 2800 years (Olasolo et al. 2016; Khademian et al. 2016), the current worldwide geothermal power capacity is about 13.3 GW (Matek 2016).

The current electricity generation from geothermal resources is restricted to a small portion of the total accessible geothermal energy, predominantly in volcanic regions with abundant groundwater (Tester et al. 2007). These geothermal reservoirs similar to oil and gas reservoirs can only be exploited until most of the contained fluid has been extracted (Olasolo et al. 2016; Khademian et al. 2017).

It is well-known that the potential of the geothermal energy at considerable depth, where the abundant heat is stored within low-permeability Hot Dry Rock (HDR) is remarkably high. In the early 1970s, a group of scientists at Los Alamos National Laboratory conducted a field-scale experiment at the Fenton Hill site with the aim of developing a technique for harvesting heat from the HDR to generate electricity. The concept of Enhanced/Engineered Geothermal System (EGS) has since then been developed and several worldwide projects move the EGS technology forward to the point where energy extraction from the HDR is ever closer to being feasible technologically and economically.

The initial concept of EGS to create a man-made reservoir with a sustained rate of heat extraction and long life was straightforward: drilling a well into the hot rock, hydraulically fracturing the rock in multiple stages and creating multiple heat exchange surfaces and ultimately drilling the production well to intercept the generated fractures. By circulating water through the system, heat can be extracted from the hot rock to the surface. The initial concept of EGS instigated a launch base for mathematically modeling thermo-hydro processes within the EGS reservoir.

Since the 1960s, several flow models have been developed to handle transport processes of subsurface flow and fracture-matrix interaction. Effective Continuum methods (ECM), Multiple Interacting Continua (MINC), the explicit discrete-fracture approach, and Discrete Fracture Network (DFN) are commonly used in hydrogeology, petroleum and geothermal for simulating flow and transport processes (Bear, Tsang, and Marsily 1993).

The ECM flow model is the most well-known flow model. It treats fractured rock as a single continuum with a set of effective properties, merging the fracture network and the rock matrix. Due to the simplicity of the ECM flow model in terms of field data requirement and computational efficiency, this flow model is used widely (e.g., Pruess et al. 1988; Nitao 1989; Berkowitz et al. 1988; Kool & Wu 1991; Wu et al. 1996).

The Multiple Interacting Continua (MINC) method covers a variety of flow scenarios. The double-porosity model proposed by Barenblatt et al., (1960) and (Warren and Root 1963), and The double-porosity model proposed by Blaskovich et al., (1983) are the most commonly used MINC methods for the practical simulation of fractured systems. Several other modified forms of the MINC methods have been developed for a better approximation of the flow and transport processes within the fractured reservoir (Kazemi 1969; Kazemi, Gilman, and Elsharkawy 1992; Pruess 1991; Pruess and Narasimhan 1985; Y.-S. Wu and Pruess 1988).

The explicit discrete-fracture approach has been developed with the aim of developing a more realistic flow model for fractured reservoir (e.g., Snow 1965; Stothoff & Or 2000). This flow model is computationally expensive and requires detailed information on all fractures embedded in the matrix. However, for modeling a smaller number of explicit fractures, such as fault zones (Bundschuh and Suárez Arriaga 2010; Khademian et al. 2012; Poeck et al. 2016), this method can provide meaningful results.

The DFM methods were developed in the late 1970s as an alternative approach to the Continuum methods. This flow model also requires detailed knowledge of fracture and matrix geometric properties and despite the recent advances in characterization methodology and numerical simulation, the use of this method for field scale problems is still conceptually challenging and computationally demanding. Still, several DFM methods have been applied (Baca, Arnett, and Langford 1984; Juanes, Samper, and Molinero 2002; Karimi-Fard, Durlofsky, and Aziz 2004; Sarda et al. 2002; Snow 1965; Sudicky and McLaren 1998).

All above-mentioned approaches can be adopted for modeling thermo-hydro processes within the EGS reservoir. The question is which method is the best fit for the EGS? Since the mathematical model of the flow and transport processes is based on the conceptual model, in complex systems such as an EGS, developing a conceptual model that can properly incorporate the geometrical details of the system is essential.

The basic assumptions of the initial concept of EGS are: (1) the flow and heat exchange occur only within the hydraulic fractures and (2) geometric properties of the incorporated domains remain unchanged during heat extraction. By adopting the initial concept of the EGS and by assuming that all information on hydraulic fractures is available (e.g., location, geometric properties, hydraulic properties), the explicit discrete-fracture approach can be applied to assess the long-term thermo-hydro processes of the idealized commercial-sized EGS. However, it is well-known that the EGS is an evolving system. In fact, circulating cold fluid through the primary hydraulic fractures induces a temperature gradient within the rock matrix, resulting in thermo-elastic stresses that propagate thermal fractures. Initially, a thermally shocked region comprising of a network of small, disorganized closely-spaced thermal cracks are formed adjacent to the hydraulic fractures. As time elapses, these small thermal cracks tend to coalesce and better-defined planar thermal fractures propagate into the rock matrix. Although thermally-induced fractures may not contribute much to the global flow and transport processes, they may provide additional locally-connected flow paths between hydraulic fractures and the rock matrix, which eventually affects fracture-matrix interactions and ultimately the reservoir performance.

The initial concept of an EGS does not incorporate thermally-induced fractures physically into reservoir simulation. Therefore, the corresponding flow and transport model cannot define the flow and heat transport within thermally-induced fractures of the reservoir.

Our proposed modified conceptual model of EGS defines the reservoir by utilizing a multi-subdomain flow model; one subdomain comprising the hydraulic fractures, one subdomain with a network of numerous smaller thermal cracks (thermally-shocked region), and one subdomain comprising the planar thermal fractures. For the first and last of the aforementioned subdomains, the explicit discrete-fracture model can be utilized. However, for the middle, the continuum medium approach can be used by treating the numerous smaller thermal cracks as a porous medium.

The main objectives of this study are (1) to formulate the flow and heat transport process within an EGS based on both the initial and the modified concept of EGS; (2) to propose an efficient computational model capable of integrating thermally-induced fractures into the simulation; and (3) to demonstrate the effect of thermally-induced fractures in EGS performance. In particular, parametric examples are developed to investigate the effects of different parameters of the thermally-induced fractures including number, width, length and permeability on flow and heat transfer in the EGS reservoir.

## 2. THE INITIAL CONCEPT OF THE EGS

The configuration of the EGS based on the initial concept comprises of two subdomains: the rock matrix, and a set of parallel equally spaced hydraulic fractures (see Figure 1). The geometries of the fractures are assumed to be thin-cylindrical fractures of radius R and uniform aperture $d_{hf}$. The same mass inflow rate is assigned to each fracture.

Jahan Bakhsh et al.

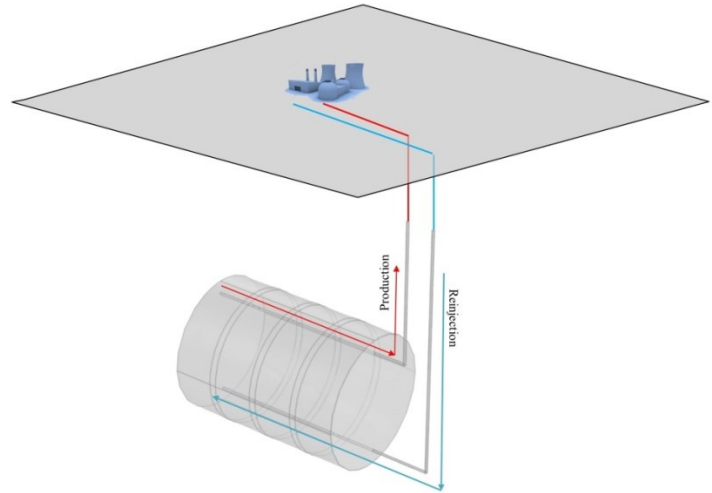

**Figure 1: Schematic of the initial concept of the EGS**

### 2.1. Mathematical formulation

A set of governing equations consisting of fluid flow and heat transfer equations must be coupled in order to model the thermo-hydro processes within the EGS reservoir mathematically. The following are additional assumptions that must be taken to set up a mathematical formulation:

- The rock matrix is assumed non-deformable, homogeneous, isotropic, and devoid of preexisting/natural fractures. Since the target area of the EGS technology is a deep intact crystalline basement rock, the properties of intact granite are used to define the rock matrix.

- Initially, the temperature is $T_\infty$ everywhere in the system and cold fluid with the constant temperature of $T_{in}$ is injected into the reservoir through the injection well.

- The fluid is assumed to be a single-phase flow within the reservoir.

- Non-linearity in the properties of the materials is not allowed and the gravitational effects are neglected.

Due to the symmetry in the $x$ and $y$ directions, the simulation domain can be restricted to half of a single fracture and two blocks of the rock matrix of width $D$ (Figure 2).

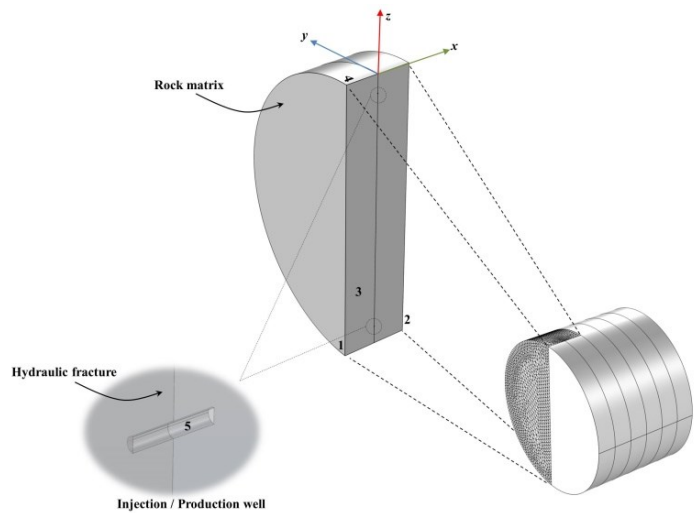

**Figure 2: Schematic of model domain used in the numerical simulation**

## 2.2. Fluid flow in reservoir

In a deep EGS reservoir, the aspect ratio of radius to aperture of a hydraulically-induced fracture is on the order of 1000 to 10000 (Armstead and Tester 1987). Accordingly, the fluid velocity across the fracture, in the $x$ direction, can be assumed constant. Following this simplification the conservation of mass of the fracture fluid can be written as follows:

$$\frac{\partial u_y}{\partial y} + \frac{\partial u_z}{\partial z} = 0 \tag{1}$$

where $u_y$ and $u_z$ are $y$ and $z$ component of fluid velocity (m/s) with respect to position. Due to the low Reynolds number, the flow in the fracture is in the laminar regime and therefore it is reasonable to utilize two-dimensional Darcy's law to govern the fluid momentum.

$$-\frac{\partial P}{\partial y} = \frac{\mu}{\kappa_{hf}} u_y \tag{2}$$

$$-\frac{\partial P}{\partial z} = \frac{\mu}{\kappa_{hf}} u_z \tag{3}$$

where $P$, $\mu$, $\kappa_{hf}$ are fluid pressure within the reservoir (MPa), the dynamic viscosity of the fluid (kg/m.s), and permeability of the hydraulic fracture (m²) respectively. Initially, the fluid velocity everywhere in the fracture subdomain is assumed zero and it is assumed that the same amount of fluid sweeps each primary hydraulic fracture.

$$m_{hf} = \frac{M_t}{N_{hf}} \tag{4}$$

where $N_{hf}$ and $M_t$ are the number of fractures and total flow rate (kg/s) within the reservoir. Accordingly, at the boundary 5 (Figure 2).

$$|\mathbf{u}| = 0.5\, m_{hf}/(2\pi r l) \tag{5}$$

The factor 0.5 accounts for the fact that half of the fracture is simulated due to the symmetry in the y direction.

## 2.3. Heat exchange in the reservoir

In EGS reservoir, fluid plays the role of energy carrier and rock matrix represents heat source. It is assumed that the heat flow within the rock matrix is via conduction. The Fourier's law is utilized to govern heat transport within the rock matrix.

$$\frac{\partial}{\partial t}\left(\rho_r C_{P_r} T_r\right) - \nabla \cdot (k_r \nabla T_r) = 0 \tag{6}$$

where $\rho_r C_{p_r}$, $T_r$, $k_r$ are volumetric heat capacity of the rock matrix at constant pressure (J/m³°C), rock matrix temperature (°C), and thermal conductivity of the rock matrix (J/m s °C). The initial temperature of the reservoir follows the geothermal gradient of the Earth.

$$T_r(x, y, z, 0) = T_\infty \tag{7}$$

where $T_\infty$ is the far field rock matrix temperature.

$$T_\infty = T_s + \mathcal{G} \cdot z \tag{8}$$

where $\mathcal{G}$, $T_s$, $z$ are the geothermal gradient (°C/km), surface temperature (°C), and length in the $z$ direction (m), respectively. It is assumed that the heat stored within the rock matrix is the only energy source that fracture fluid can take away from the rock. Following this assumption, symmetry boundary at surface 1, 2, and 3 and the no-flux boundary condition at surface 4 are applied (Figure 2).

$$-\mathbf{n} \cdot \nabla T_r = 0 \tag{9}$$

The interaction between the rock matrix and the fracture flow in terms of heat exchange can be defined as follows:

$$q = -2 \times \left[\frac{\partial T_r(0, y, z, t)}{\partial x}\right] \tag{10}$$

.where q is the heat flux from rock matrix toward hydraulic fracture from two faces of the fracture. The energy balance equation within the fluid domain can be written as follows:

$$\rho_f C_{p_f} \frac{\partial T}{\partial t} + \rho_f C_{p_f} \mathbf{u} \cdot \nabla T = \nabla \cdot (k_f \nabla T) + Q \tag{11}$$

Jahan Bakhsh et al.

where $\rho_f C_{p_f}$, **u**, $k_f$, are volumetric heat capacity of the fluid (J/m³°C), fluid velocity (m/s), thermal conductivity of the fluid (J/m s °C).

$$Q \approx q = -2 \times \left[\frac{\partial T_r(0,y,z,t)}{\partial x}\right] \quad (12)$$

By substituting $Q$ into the Eq.11 into the energy balance this equation can be rewritten as follows:

$$\frac{\partial T}{\partial t} + \frac{\partial(Tu_y)}{\partial y} + \frac{\partial(Tu_z)}{\partial z} = \frac{k_f}{\rho_f C_{p_f}}\left(\frac{\partial^2 T}{\partial y^2} + \frac{\partial^2 T}{\partial z^2}\right) + \frac{q}{\rho_f C_{p_f} d_{hf}} \quad (13)$$

Fluid with a lower temperature than its surrounding rock matrix is injected into the reservoir, therefore at boundary 5.

$$T(t) = T_r(t) = T_{in} < \bar{T}_\infty \quad (14)$$

Where, $T_{in}$ is the injection temperature(°C).

**2.4. Numerical implementation**

In EGS which is a system with geometrical peculiarity, discretizing subdomains explicitly creates an enormous number of fine elements for the fractures and thousands of coarse elements for the rock matrix. This combination of elements amasses a tremendous computational burden. There are several approaches available in the literature that can tackle this problem and reduce the computation time (see, e.g., Pinder et al. 1993).

In this study, the EGS reservoir is treated as a multi-dimensional system consisting of two subdomains of different spatial dimensions; (1) a three-dimensional rock matrix, and (2) two-dimensional hydraulic fractures. Adopting this approach eliminates the need to create fractures with a high aspect ratio by considering hydraulic fractures as 2D interior boundaries of the 3D rock matrix. This approximation allows us to discretize the rock matrix, and the fractures using tetrahedral (3D) and by triangular (2D) elements respectively.

COMSOL Multiphysics is selected as a framework for numerical implementation. The conservation equations are discretized spatially by employing the finite-element method.

**3. THE MODIFIED CONCEPT OF THE EGS**

The scheme of the EGS, based on the modified concept, is divided into four subdomains (Figure 3):

(1) The rock matrix, representing the heat source;

(2) The hydraulic fractures, representing the *globally connected* flow paths and heat exchange surfaces

(3) The planar thermal fractures, representing, *locally connected* flow paths and heat exchange surfaces

(4) The thermally-shocked region, representing a *buffer zone* with high storage capacity, connecting the hydraulic fractures, the rock matrix, and the planar thermal fractures

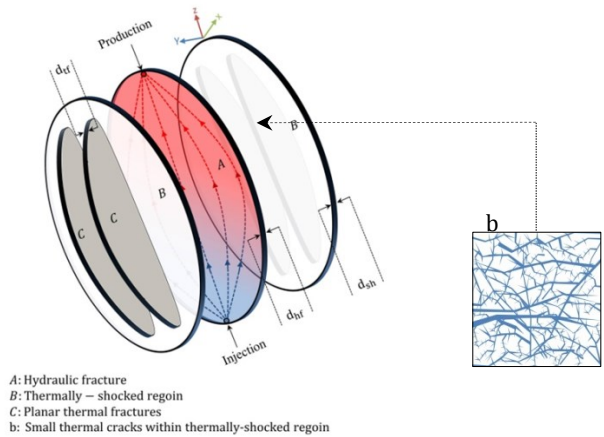

A: Hydraulic fracture
B: Thermally – shocked regoin
C: Planar thermal fractures
b: Small thermal cracks within thermally-shocked regoin

**Figure 3: The scheme of the modified concept of the EGS**

Jahan Bakhsh et al.

Since the modified concept of EGS is considered as a dynamic system, the incorporated subdomains must be integrated into the simulation chronologically by following the sequence of events in the EGS system. In the EGS reservoir, the sequence of events can be defined chronologically as follows: initially, when the cold fluid sweeps the hydraulic fractures, a thermally-shocked region comprised of a network of small, disorganized thermal cracks form in the rock adjacent to the hydraulic fractures. As time elapses, these small thermal cracks tend to coalesce and form better-defined planar thermal fractures perpendicular to the hydraulic fractures. Following the chronological order of the events in an EGS system, the model components can be integrated into the simulation sequentially.

**3.1. Mathematical formulation**

A hybrid model comprised of the explicit discrete-fracture and the Effective Continuum Models can be utilized to describe the flow and heat transport within the reservoir. There are a few assumptions for the thermally-shocked region and the planar thermal fractures explained below.

➢ *Thermally-shocked region*

The geometry of the thermally-shocked regions is assumed to be identical, i.e., a thin-cylindrical slab of radius $R$ with a uniform width of $d_{sh}$ adjacent to the hydraulic fractures. The thermally-shocked region is idealized as a porous medium, located on both sides of the hydraulic fracture resembling a thin-cylindrical slab of radius $R$ with a uniform width of $d_{sh}$. A Gaussian distribution is used to define permeability distributions across the thermally-shocked region. As shown in Figure 4 it is assumed that permeability and porosity of the thermally shocked region are variable spatially and temporally.

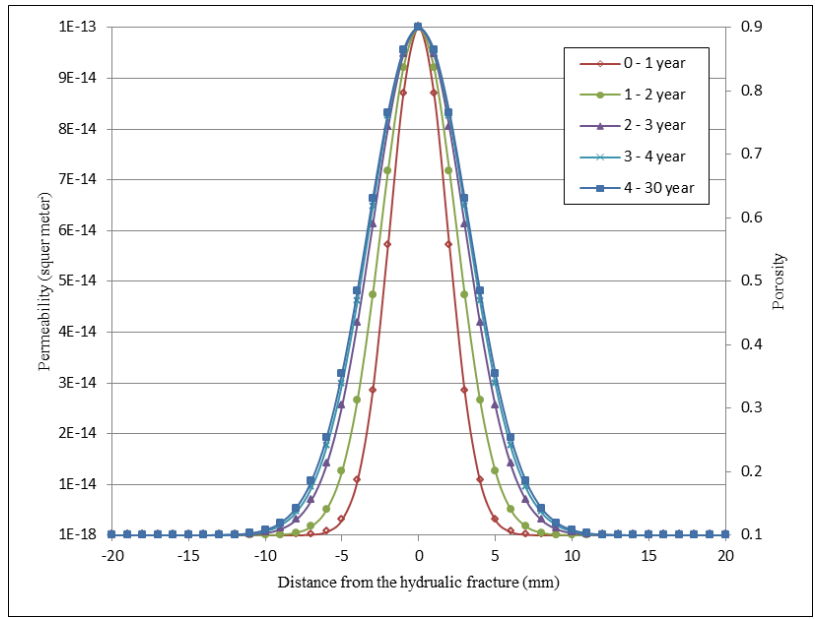

**Figure 4: The assigned distribution of the porosity and permeability across the thermally-shocked region.**

➢ *Planar thermal fractures*

The geometry of the planar thermal fractures is assumed to be identical, i.e., thin planar fractures of length $l_{th}$ and uniform aperture of $d_{th}$. It is also assumed that the planar thermal fractures are extended in the *x* direction, into the rock matrix equally from both walls of the hydraulic fractures.

The governing equations for all subdomains are provided in Table 1.

Jahan Bakhsh et al.

Table 1: The governing equations for solving fluid flow and heat transfer within each subdomain numerically

| Subdomain | Fluid flow governing equations | Heat transfer governing equations |
|---|---|---|
| Rock matrix | $\frac{\partial}{\partial t}(\varepsilon_r \rho_r) + \nabla \cdot (\rho_r u) = 0$ <br> $u = -\frac{\kappa_r}{\mu} \nabla p$ | $(\rho C_p)_r^{eff} \frac{\partial T}{\partial t} + \rho_f C_{p_f} u \cdot \nabla T + \nabla \cdot q = 0$ <br> $q = -k_r^{eff} \nabla T$ |
| Hydraulic fractures | $d_{hf} \frac{\partial}{\partial t}(\varepsilon_{hf} \rho_f) + \nabla_t \cdot (d_{hf} \rho_f u) = 0$ <br> $u = -\frac{\kappa_{hf}}{\mu} \nabla_t p$ | $d_{hf}(\rho C_p)_{hf}^{eff} \frac{\partial T}{\partial t} + d_{hf} \rho_f C_{p_f} u \cdot \nabla_t T + \nabla_t \cdot q_{hf} = n \cdot q$ <br> $q_{hf} = -d_f k_{hf}^{eff} \nabla_t T$ |
| Thermally-shocked region | $d_{sh} \frac{\partial}{\partial t}(\varepsilon_{sh} \rho_f) + \nabla_t \cdot (d_{sh} \rho_f u) = 0$ <br> $u = -\frac{\kappa_{sh}}{\mu} \nabla_t p$ | $d_{sh}(\rho C_p)_{sh}^{eff} \frac{\partial T}{\partial t} + d_{sh} \rho_f C_{p_f} u \cdot \nabla_t T + \nabla_t \cdot q_{sh} = n \cdot q$ <br> $q_{sh} = -d_{sh} k_{sh}^{eff} \nabla_t T$ |
| Planar Thermal Fractures | $d_{tf} \frac{\partial}{\partial t}(\varepsilon_{hf} \rho_f) + \nabla_t \cdot (d_{tf} \rho_f u) = 0$ <br> $u = -\frac{\kappa_{tf}}{\mu} \nabla_t p$ | $d_{tf}(\rho C_p)_{tf}^{eff} \frac{\partial T}{\partial t} + d_{tf} \rho_f C_{p_f} u \cdot \nabla_t T + \nabla_t \cdot q_{tf} = n \cdot q$ <br> $q_{tf} = -d_{tf} k_{tf}^{eff} \nabla_t T$ |

### 3.2. Numerical implementation

In order to pair both the hydraulic fractures and the planar thermal fractures with the rock matrix, the explicit discrete-fracture approach can be adopted. For the thermally-shocked region, the explicit discrete-fracture model cannot be utilized for integrating this region into the simulation. There are two reasons that make the explicit discrete-fracture model inapplicable for integrating the thermally-shocked region:

✓ First, it is assumed that the properties of the thermally-shocked region evolve in time and space, therefore, the dimensional reduction offered by the explicit discrete-fracture approach, cannot capture the properties of this region.

✓ Second, defining the thermally-shocked region as an interior boundary leads to placing an interior boundary adjacent to another interior boundary, i.e., the hydraulic fracture. The adjacent boundaries cannot be defined by an explicit discrete-fracture approach.

Our proposed hybrid approach is capable of integrating all four subdomains efficiently. In this approach, the concept of an Effective Continuum Method (ECM) is used to summarize the hydraulic fracture and thermally-shocked region into an equivalent subdomain with a set of "effective" parameters (Figure 5). The equivalent subdomain which contains enough information for both the hydraulic fracture and thermally-shocked region can be integrated into the simulation by representing the thermally-shocked\hydraulic fracture system as a 2D interior boundary.

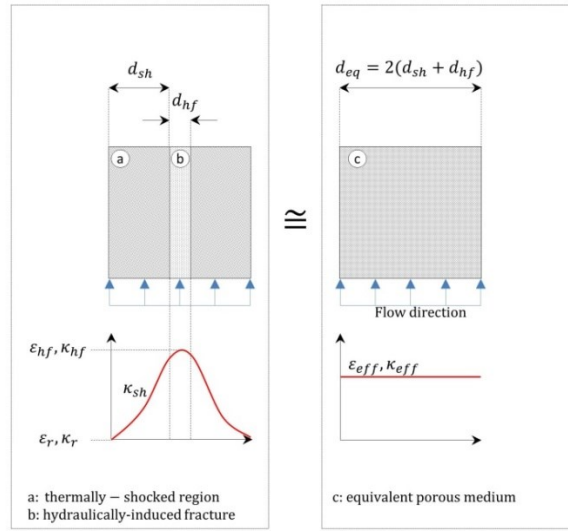

**Figure 5: The schematic of summarizing the hydraulically-induced fracture and thermally-shocked region into the equivalent subdomain with a set of "effective" parameters by utilizing the Effective Continuum Method (ECM)**

## 4. MODEL SETUP; CONFIGURATION AND ASSUMPTIONS

In this work, based on the initial concept, an idealized EGS reservoir, comprising of a doublet system and a set of 12 vertically oriented hydraulic fractures at an average depth of 5 km is considered as a basic case (CS-A). Table 2 and Table 3 provide geometrical and site-specific properties of the reservoir and all other fixed parameters used in the numerical simulation.

**Table 2: The geometrical and site-specific properties of the reservoir**

| Parameter | Symbol | Value/Unit |
|---|---|---|
| Average reservoir depth | $\bar{Z}$ | 5 km |
| Thermal gradient | $\mathcal{G}$ | 60 °C/km |
| Average reservoir temperature | $\bar{T}_\infty$ | 315 °C/km |
| Hydraulic fracture radius | $R$ | 500 m |
| Half fracture separation distance | $D$ | 60 m |
| Number of Hydraulic fractures | $N_{hf}$ | 12 |
| Hydraulic fracture aperture | $d_{hf}$ | 4 mm |
| Hydraulic fracture porosity | $\varepsilon_{hf}$ | 0.9 |
| Hydraulic fracture permeability | $\kappa_{hf}$ | 1E-13 |
| Total injection rate | $M_t$ | 100 kg/s |
| Reservoir life time | $t$ | 30 years |

Table 3: Fixed input parameters used in the numerical simulation

| Parameter | Symbol | Value/Unit |
|---|---|---|
| Hot Rock | | |
| Thermal conductivity | $k_r$ | 2.9 W/(m K) |
| Heat capacity | $C_{p_r}$ | 850 J/(kg K) |
| Density | $\rho_r$ | 2600 kg/m$^3$ |
| Porosity | $\varepsilon_r$ | 0.1 |
| Permeability | $\kappa_r$ | 1E-18 m$^2$ |
| Fluid (water) | | |
| Thermal conductivity | $k_f$ | 0.6 W/(m K) |
| Heat capacity | $C_{p_f}$ | 4200 J/(kg K) |
| Density | $\rho_f$ | 977.8 kg/m$^3$ |
| Dynamic viscosity | $\mu$ | 1E-3 Pa.s |
| Reservoir inflow Temperature | $T_{in}$ | 50 °C |

The modified concept of the EGS is utilized to set up two more cases (i.e., CS-B and CS-C). For the CS-B case, the reservoir comprises of a hydraulic fracture and four planar thermal fractures perpendicular to the hydraulic fracture. The properties of planar thermal fractures for the CS-B case are listed in Table 4. It is also assumed that the planar thermal fractures are hydraulically inactive during the first five years of the heat extraction, so their contribution to the flow and heat transport process is negligible until the fifth year.

In the CS-C case, in addition to a hydraulic fracture and four planar thermal fractures, two thermally-shocked regions of width 0.02 m on each side of the hydraulic fracture are included. The porosity and permeability of the thermally-shocked region are obtained following the proposed Gaussian function (Figure 4).

Table 4: The properties of the planar thermal fractures

| Parameter | Symbol | Value/Unit |
|---|---|---|
| Number of fractures | $N_{tf}$ | 4 |
| Aperture | $d_{tf}$ | 1 mm |
| Penetration length | $l_{tf}$ | 60 m |
| Porosity | $\varepsilon_{tf}$ | 0.9 |
| Permeability | $\kappa_{tf}$ | 1E-14 m$^2$ |

## 5. NUMERICAL SIMULATION RESULTS

A comparative study between CS-A, CS-B, and CS-C cases reveals the capability of the developed hybrid-dimensional model for integrating thermally-induced fractures into the reservoir simulation. The effects of (1) the planar thermal fractures and (2) the thermally-shocked region on the reservoir performance are examined by comparing the CS-A case with CS-B and CS-C cases respectively. Figure 6 shows the evolution of a dimensionless temperature profile in the rock matrix during 30 years of production for all three cases.

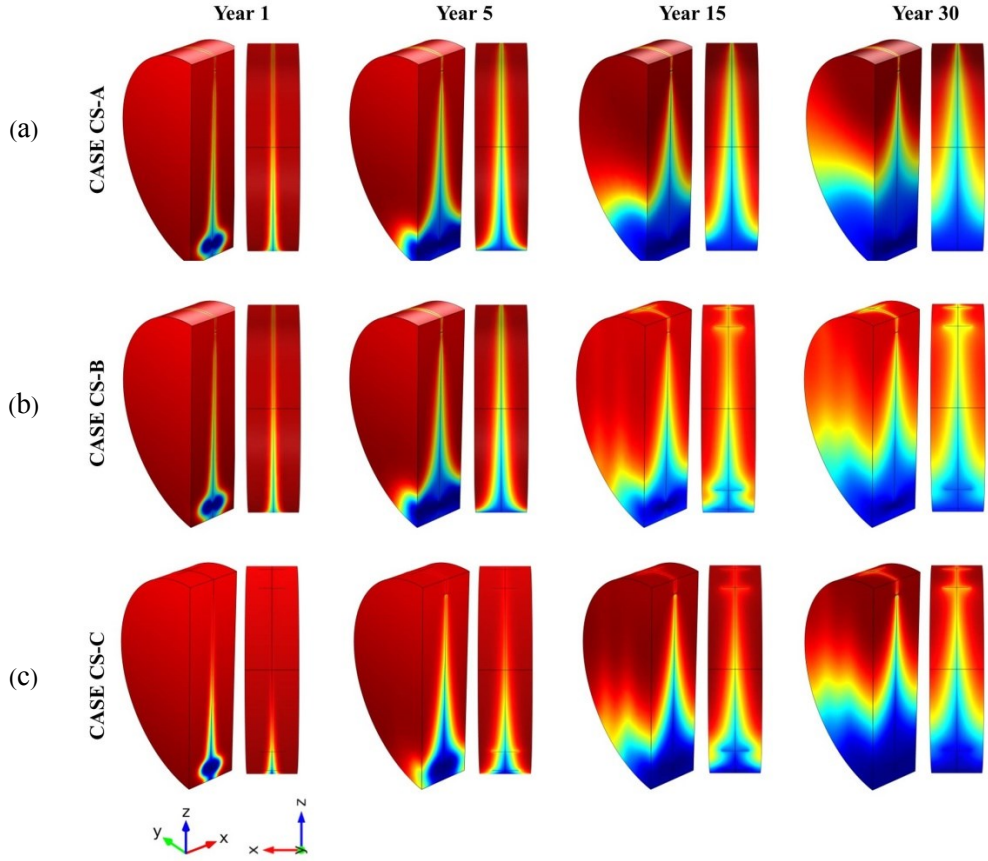

**Figure 6: The evolution of dimensionless temperature profile of the rock matrix during 30 years of production for (a) CS-A, (b) CS-B, and (c) CS-C cases**

For the CS-A case the injection of cold fluid into the hot reservoir induces a horizontal temperature gradient. The intensity of the induced temperature gradient is highest near the injection point, resulting in a localized convex shape for the cold front. As shown in the Figure 6a as time elapses, the isotherms in the rock matrix advance into the rock matrix following the convex contour close to the injection point and become parallel to the fracture near the production point.

In the CS-B case (Figure 6b), the presence of the planar thermal fractures provides additional flow paths and heat exchange surfaces to the model. Since during the first five years of heat extraction, the planar thermal fractures are assumed hydraulically inactive, the advancement of the cold front into the rock matrix for the CS-B case is similar to the CS-A case. However, after year five as soon as the planar thermal fractures start contributing to the thermo-hydro processes, the advancement of the cold front into the rock matrix is affected. The temperature profile of the CS-B case, at year 15 and 30, shows that although the cold front exhibits a convex shape close to the injection point, it also extends along the planar thermal fractures.

For the CS-C case (Figure 6c), the presence of the thermally-shocked region on both sides of the hydraulic fracture introduces a porous subdomain with high storage capacity to the model. Fig.9c shows that for the CS-C case, during the first five years of the heat extraction when the thermally-shocked region is evolving, the advancement of the cold front is decelerated compared to the advancement of the cold front in both CS-A and CS-B cases. This is because the fluid can permeate into the thermally-shocked region, which would increase the exposure time of the fluid to the hot rock. Similar to case CS-B case, as time elapses, the irregularity in the cold front caused by the planar thermal fracture can be clearly observed.

Figure 7 shows the non-dimensional outlet temperature over 30 years of heat extraction for CS-A, CS-B, and CS-C cases. During the first five years of operation, in which the planar thermal fractures are inactive hydraulically, the temperature drawdown rate is faster than throughout the rest of the operation time for all three cases. For the CS-A case, the dimensionless temperature at the production well drops to 0.5 after 30 years, however, for CS-B and CS-C cases the production temperature declines to 0.55 and 0.62 respectively. Although the overall trend for all cases is the same, for the CS-B case as soon as the planar thermal fractures become hydraulically-active, there is a sudden increase in the production temperature which reverses the temperature drawdown for couple years. This is because the contribution of the planar thermal fractures in the thermos-hydro process is started at year 5. For CS-C case, the presence of Jahan Bakhsh et al.

the thermally-shocked region slows the temperature drawdown. The slight increase in production temperature at year 5, due to the presence of the planar thermal fracture can also be observed for the CS-C case; however, the effect of the thermally shocked region is dominant. The presence of the planar thermal fractures increases the production temperature by 6% on average over 30 years of production; whereas, the combination of planar thermal fractures and a thermally-shocked region increases the production temperature by 24% on average over the operation time.

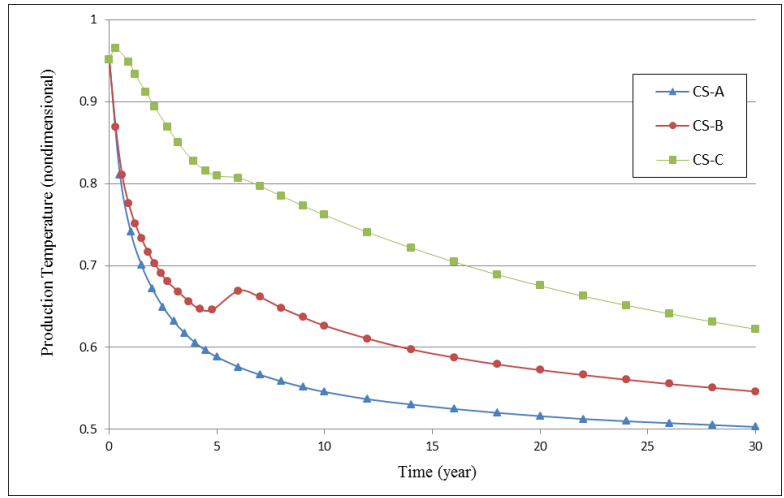

**Figure 7: The dimensional outlet temperature over 30 years of heat extraction for CS-A, CS-B, and CS-C cases**

**5.1. Sensitivity analysis of heat extraction to the planar thermal fracture parameters**

The effects of the different parameters of the planar thermal fracture- including the number, the aperture, the length and the permeability- on the reservoir performance are examined in this section. The example scenarios with and without thermally-shocked regions are listed in Table 5.

**Table 5: The case scenario examples with/without thermally-shocked**

| CASE | Planar thermal fracture | | | | Thermally-shocked region |
| --- | --- | --- | --- | --- | --- |
| | Penetration length | aperture | permeability | Number | |
| | $l_{tf}$ (m) | $d_{tf}$ (mm) | $\kappa_{tf}$ ($10^{-14}$ m$^2$) | $N_{tf}$ | |
| CS1 | 40,60,80 | 1 | 1 | 4 | Not Considered |
| CS2 | 60 | 0.5,1,2 | 1 | 4 | |
| CS3 | 60 | 1 | 0.1, 1, 10 | 4 | |
| CS4 | 60 | 1 | 1 | 2,4,6 | |
| CS5 | 40,60,80 | 1 | 1 | 4 | Considered |
| CS6 | 60 | 0.5,1,2 | 1 | 4 | |
| CS7 | 60 | 1 | 0.1, 1, 10 | 4 | |
| CS8 | 60 | 1 | 1 | 2,4,6 | |

5.1.1 Effect of penetration length

Figure 8 shows dimensionless production temperature during the operation time for cases CS1 and CS5. There are two clusters of curves; the lower cluster (solid curves) represents the CS1 case in which the thermally-shocked region is not included physically in the simulation, and the upper cluster (dashed curves) represents the CS5 case in which the planar thermal fractures and the thermally-shocked region are included. As shown in Figure 8, for both cases, by increasing the penetration length the production temperature drawdown is decelerated because deeper penetration length provides larger heat exchange surface area and allows the fluid to access a larger volume of the hot rock. For the CS1 case (solid curves), by increasing the penetration length a sudden increase in the production temperature at year five becomes more pronounced and finally, the curves converge at the end of operation time. On the other hand, for CS5 case (dashed curves), increasing the penetration length stops the temperature drawdown temporarily after year 5, and then drawdown continues steadily until the end of operation time. In other words, the presence of a thermally-shocked region distributes the effect of the penetration length throughout the operation time and the production temperature drops steadily during the operation years. In general, the effect of the penetration length of the planar thermal fracture on reservoir performance is negligible (less than 1%).

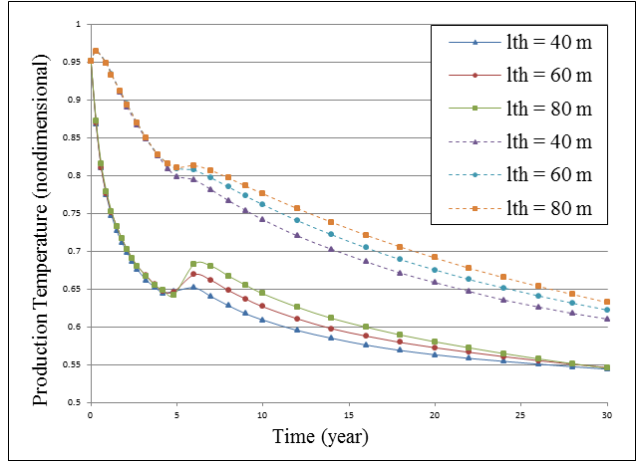

**Figure 8: The dimensional outlet temperature over 30 years of heat extraction for CS1 (solid curves) and CS5 (dashed curves) cases**

5.1.2. Effect of the aperture of the planar thermal fractures

Figure 9 shows the dependence of dimensionless production temperature on the aperture of the planar thermal fractures during the operation time, using the parameters listed for cases CS2 and CS6 in Table 5. As seen in Figure 9, the production temperature is insensitive to the fracture aperture for both CS2 (solid curves) and CS6 (dashed curves) cases. Since the planar thermal fractures are oriented away from the global flow, the average velocity within planar thermal fractures is significantly smaller than the average velocity within the hydraulic fracture. Therefore, the aperture of the planar thermal fracture has little influence on the velocity field of this subdomain and consequently, an insignificant effect on the fluid temperature carried out of the reservoir. It can be concluded that effect of the aperture of planar thermal fracture on reservoir performance is negligible (less than 1%).

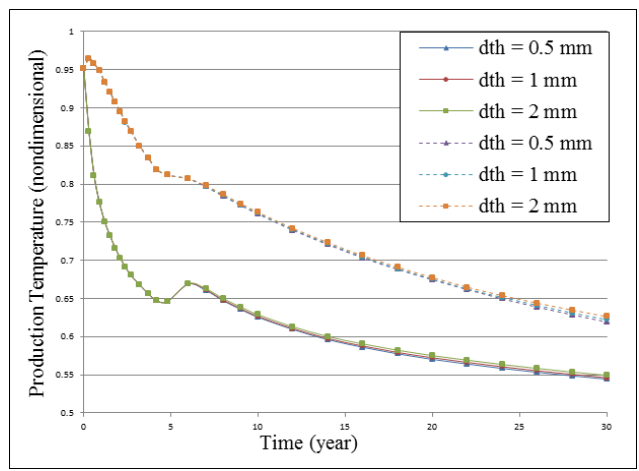

**Figure 9: The dimensional outlet temperature over 30 years of heat extraction for CS2 (solid curves) and CS6 (dashed curves) cases**

5.1.3 Effect of the permeability of the planar thermal fractures

Figure 10 shows the dependence of the dimensionless production temperature on the permeability of the planar thermal fractures during the operation time, using the parameters listed for cases CS3 and CS7 in Table 5. As shown in Figure 10 increasing the permeability of the planar thermal fracture one order of magnitude from 1E-15 to 1E-14 m2 slows the production temperature drawdown slightly for both CS3 and CS7 cases. Although the same behavior is expected when the permeability is increased further, the temperature curve for the cases with the permeability of 1E-13 m2 drops to the lower value than cases with the permeability of 1E-14 and 1E-15 m$^2$. In general, the effect of the permeability of the planar thermal fracture on reservoir performance is negligible (less than 1%).

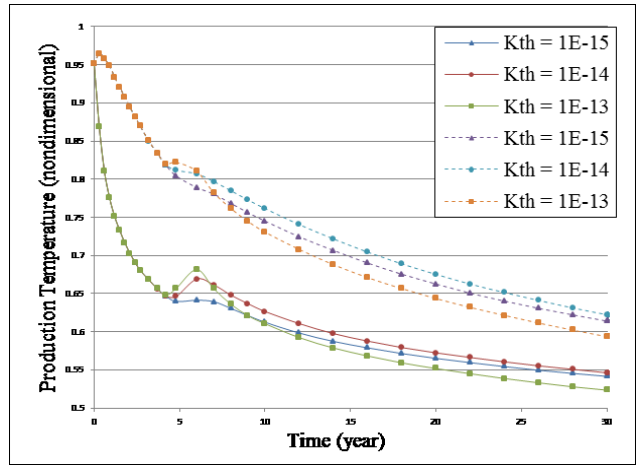

**Figure 10: The dimensional outlet temperature over 30 years of heat extraction for CS3 (solid curves) and CS7 (dashed curves) cases**

5.1.4 Effect of the number of planar thermal fracture

Figure 11 shows the dependence of the dimensionless production temperature on the number of planar thermal fractures during the operation time, using the parameters listed for cases CS4 and CS8 in Table 5. As seen in Fig, for both cases, increasing the number of planar thermal fractures slows the production temperature drawdown because a larger number of planar thermal fractures provide larger heat exchange surface area and allows the fluid to access a larger volume of the hot rock. The results indicate that although the planar thermal fractures represent locally connected flow paths and heat exchange surfaces, a large number of them is desirable for mitigating the thermal drawdown of a reservoir. Results show that increasing the number of thermal planar fractures from two to six fractures increases the production temperature by 5% on average over the operation time.

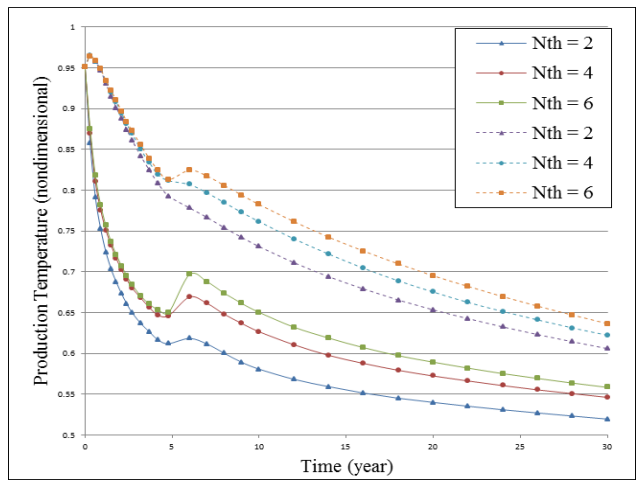

**Figure 11: The dimensional outlet temperature over 30 years of heat extraction for CS4 (solid curves) and CS8 (dashed curves) cases**

## 6. CONCLUSION

Although thermally-induced fractures are believed to improve EGS performance, they are usually excluded from the EGS reservoir simulation. In this work, we introduce the modified concept of the EGS, in which the physical model of the EGS reservoir comprises four subdomains: the hot rock matrix, the hydraulic fractures, the thermally-shocked region and the planar thermal fractures. A hybrid

Jahan Bakhsh et al.

flow model capable of integrating thermally induced features into the EGS reservoir simulation is developed. COMSOL Multiphysics is utilized as a finite element framework and three numerical examples are solved based on the initial and modified concepts of the EGS. The effects of the thermally-induced fractures on the thermal performance of an EGS are quantified. Finally, parametric scenarios are developed to investigate the sensitivity of the thermal performance in the EGS on the properties of the planar thermal fractures. Based on the study results, we find:

(1) While it is currently assumed that in the presence of a well-connected, large-scale hydraulic fracture, the locally-connected, small-scale thermally-induced fractures do not contribute significantly to the global flow and transport within the EGS reservoir; in this work, we show that thermally-induced fractures can affect the thermal performance of an EGS by providing additional connection area for interflow between the rock matrix and the well-connected, large-scale hydraulic fracture. The results of the numerical example show that integrating thermally-induced fractures into the simulation enhance the thermal performance of the EGS reservoir up to 24% on average over 30 years of production.

(2) The results of the parametric examples show that the penetration length, the aperture, and the permeability of the planar thermal fractures have a negligible effect on the thermal performance of the EGS. Hence, increasing the number of thermal planar fractures from two to six fractures increases the production temperature by 5% on average over the operation time.

(3) The results confirm that adopting the initial concept of the EGS underestimates the performance of EGS technology. Hence, although it is simplified, the proposed hybrid flow model (developed based on a modified conceptual model) evaluates the EGS performance more realistically, which ultimately enhances the overall competitiveness of EGS technology as a form of green energy.

Jahan Bakhsh et al.